\documentclass[twocolumn,superscriptaddress,floatfix,prl]{revtex4-1}
\usepackage{graphicx,amsfonts,amssymb,amsmath,hyperref,hypcap,enumerate}
\usepackage{color}

\newcommand{\ba}{\boldsymbol{a}}

\newcommand{\RR}{\boldsymbol{R}}
\newcommand{\rr}{\boldsymbol{r}}

\newcommand{\kk}{\boldsymbol{k}}
\newcommand{\dd}{\boldsymbol{d}}

\newcommand{\GG}{\boldsymbol{G}}

\begin{document}
\title{Topological insulators in twisted transition metal dichalcogenide homobilayers}

\author{Fengcheng Wu}
\affiliation{Materials Science Division, Argonne National Laboratory, Argonne, Illinois 60439, USA}
\affiliation{Condensed Matter Theory Center and Joint Quantum Institute, Department of Physics, University of Maryland, College Park, Maryland 20742, USA}

\author{Timothy Lovorn}
\affiliation{Department of Physics, University of Texas at Austin, Austin, Texas 78712, USA}

\author{Emanuel Tutuc}
\affiliation{Department of Electrical and Computer Engineering, Microelectronics Research Center, The University of Texas at Austin, Austin, Texas 78758, USA}

\author{Ivar Martin}
\affiliation{Materials Science Division, Argonne National Laboratory, Argonne, Illinois 60439, USA}

\author{A. H. MacDonald}
\affiliation{Department of Physics, University of Texas at Austin, Austin, Texas 78712, USA}


\begin{abstract}
We show that  moir\'e bands of twisted homobilayers
can be topologically nontrivial, and illustrate the tendency by studying  
valence band states in $\pm K$ valleys of twisted bilayer transition metal dichalcogenides, in particular, bilayer MoTe$_2$.  Because of the large spin-orbit splitting at the monolayer valence band maxima, 
the low energy valence states of the twisted bilayer MoTe$_2$ at $+K$ ($-K$) valley can be described using a two-band model
with a layer-pseudospin magnetic field $\boldsymbol{\Delta}(\boldsymbol{r})$ that has 
the moir\'e period.  
We show that  $\boldsymbol{\Delta}(\boldsymbol{r})$ has a 
topologically non-trivial skyrmion lattice texture
in real space, and that the topmost moir\'e valence bands
provide a realization of the Kane-Mele quantum spin-Hall model, i.e., the two-dimensional time-reversal-invariant  topological insulator.
Because the bands narrow at small twist angles, a rich set of broken symmetry insulating states can occur at integer numbers of electrons per moir\'e cell.
\end{abstract}

\maketitle
{\it Introduction.---}
Moir\'e superlattices form in van der Waals bilayers with 
small differences between the lattice constants or orientations of the individual layers,
and often dramatically alter electronic 
properties \cite{Hunt2013, Dean2013, Wang2015, Kim_bilayer, Spanton2018,Chen2018}. 
In the presence of long-period moir\'e patterns, electronic states can be described by 
continuum model Hamiltonians with the moir\'e periodicity and spinors whose dimension is equal to the 
total number of bands, summed over layers, in the energy range of interest.
Application of Bloch's theorem then
gives rises to moir\'e bands \cite{Bistritzer2011}. 
Because the moir\'e pattern often generates spatial confinement,
moir\'e bands can be narrow, enhancing the importance of electronic
correlations. The flat bands of magic-angle twisted 
bilayer graphene, in which correlated insulating and superconducting
states have been discovered \cite{Cao2018Magnetic, Cao2018Super},  provide a prominent example.
The study of moir\'e flat bands has recently become an active area of 
experimental and theoretical research centered on efforts to identify 
promising bilayer structures, and on topological characterization 
and many-body interaction physics \cite{ Wu_Hubbard2018, Po2018,  Wu_phonon2018, Naik2018ultraflat, Jung2018, Senthil2018}.

When the two layers are formed from the same material (homobilayers), 
both must be treated on an equal footing.
The $\pm K$-valley valence bands of semiconductor group-VI transition metal dichalcogenide(TMD) monolayers provide a prototypical model system because strong spin-orbit 
coupling and broken inversion symmetry lifts spin-degeneracy \cite{Di2012}, and the corresponding homobilayer can be described by a two-band model with layer pseudospins at each valley. The moir\'e pattern's 
periodic modulation can then be accounted for by a
scalar potential and a pseudo magnetic field $\boldsymbol{\Delta}(\rr)$ 
whose components are the coefficients of the layer Pauli matrix 
expansion of the two-band Hamiltonian, {\it i.e.}  
$\Delta_{x}$ and $\Delta_{y}$ are the real and imaginary parts of the interlayer tunneling
amplitude and $\Delta_{z}$ is the potential difference between layers. 
The field $\boldsymbol{\Delta}(\rr)$ inherits the moir\'e pattern periodicity
and plays a key role in the discussion below. 

\begin{figure}[t]
    \includegraphics[width=1\columnwidth]{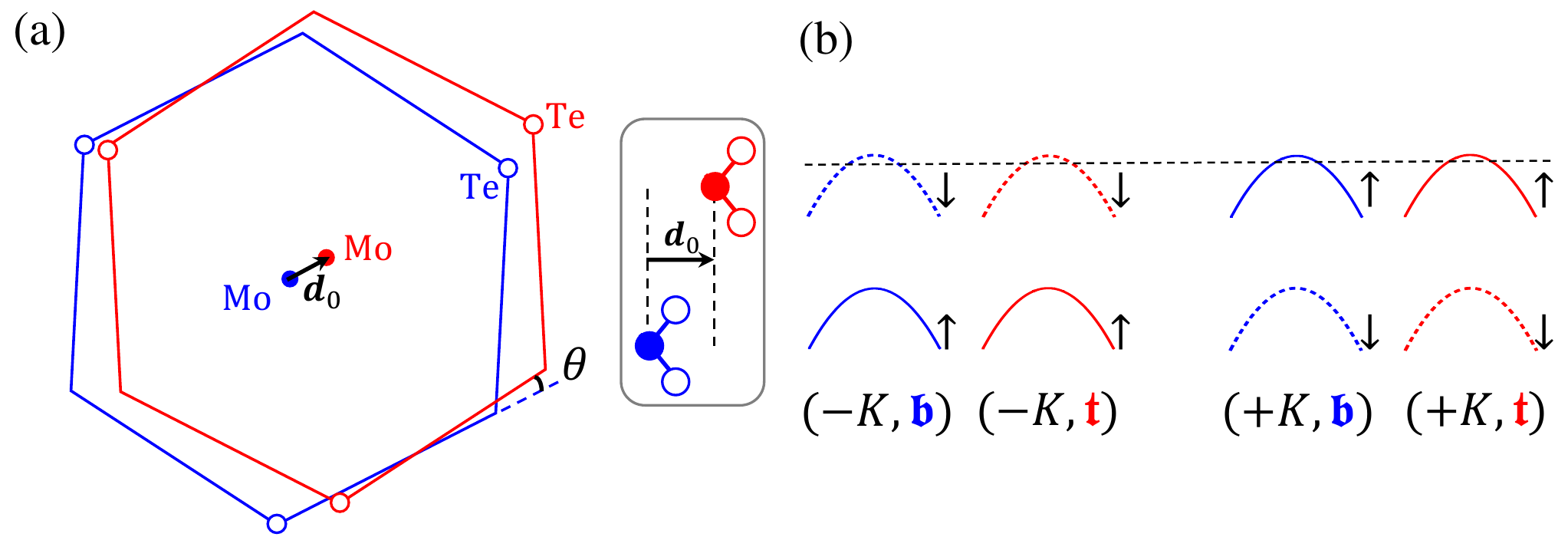}
	\caption{(a) Top view of AA stacked MoTe$_2$ homobilayer with a small twist angle $\theta$ and an in-plane displacement $\dd_0$. The inset is a schematic side view.  (b) $\pm K$ valence bands in the AA stacking case when 
	interlayer coupling is neglected.} 
	\label{Fig:Moire_Lattice}
\end{figure}

In this Letter, we focus on the MoTe$_2$ bilayer with AA stacking 
[Fig.~\ref{Fig:Moire_Lattice}], for which valence band maxima are located in $\pm K$ valleys according to our first-principles calculations as shown in the Supplemental Material(SM)\cite{SM}.  For this system, we find that 
$\boldsymbol{\Delta}(\rr)$ has a skyrmion lattice texture in real space,
and that the moir\'e bands carry valley-contrasting Chern numbers. The topological moir\'e bands can provide a realization of the Kane-Mele model, where the effective gauge potential is generated by the momentum shift between the two twisted layers.
When the bilayer is polarized by a 
vertical displacement potential, the band Chern numbers are driven 
to zero before $\boldsymbol{\Delta}(\rr)$ becomes  topologically trivial in real space. 
In partially filled topological flat bands, interactions can, for example, break time-reversal symmetry to form quantum anomalous Hall states. 

{\it Aligned bilayers---}
To derive a moir\'e continuum Hamiltonian, we start by analyzing the electronic 
structure of an aligned bilayer \cite{Jung2014}. 
Because the $\pm K$ valleys are related by time-reversal symmetry $\hat{\mathcal{T}}$, we can focus on the $+K$ valley. 
In an AA stacked TMD homobilayer [Fig.~\ref{Fig:Moire_Lattice}], the valence states at the $+K$ valley 
valence band maximum are mainly of $d_{x^2-y^2}+id_{xy}$ orbital character, have
spin up ($\uparrow$) along $\hat{z}$ axis \cite{Di2012}, and are separated from spin-down ($\downarrow$) states by strong
spin-orbit splitting.  Retaining only the spin up valence-band states at the $+K$ valley 
yields the two-band $k\cdot p$ Hamiltonian \cite{SM},
\begin{equation}
\mathcal{H}_{\uparrow}(\theta=0, \dd_0)=\begin{pmatrix}
-\frac{\hbar^2 \kk^2}{2 m^*} +\Delta_{\mathfrak{b}}(\dd_0) & \Delta_{T}(\dd_0) \\
\Delta_{T}^{\dagger}(\dd_0) & -\frac{\hbar^2 \kk^2}{2 m^*} +\Delta_{\mathfrak{t}}(\dd_0)
\end{pmatrix},
\label{Hd0}
\end{equation}
with parameters that depend on the displacement $\dd_0$ between the aligned layers.
In Eq.~(\ref{Hd0}), $\mathfrak{b}$ and $\mathfrak{t}$ refer to bottom ($\mathfrak{b}$) 
and top ($\mathfrak{t}$) layers, 
$\kk$ is momentum measured from $+K$ point, $m^*$ is the valence band effective mass that is approximately independent of $\dd_0$ \cite{SM}, 
$\Delta_{\mathfrak{b},\mathfrak{t}}$ are layer-dependent energies, 
and  $\Delta_{T}$ is an inter-layer tunneling amplitude.
The dependence of $\Delta_{\alpha}$ ($\alpha = \mathfrak{b},\mathfrak{t}, T $) on $\dd_0$ is constrained
by the symmetry properties of the bilayer.  
The two-dimensional lattice periodicity of the aligned bilayers implies that
the $\Delta_{\alpha}$ are periodic functions of $\dd_0$. 
A $z \leftrightarrow -z$ mirror operation interchanges $\mathfrak{b}$ and $\mathfrak{t}$
and maps displacement $\dd_0$ to $-\dd_0$, implying that $\Delta_{\mathfrak{t}}(\dd_0)=\Delta_{\mathfrak{b}}(-\dd_0)$. 
Threefold rotation around the $\hat{z}$ axis requires that $\Delta_{\mathfrak{b}}$ and 
$\Delta_{\mathfrak{t}}$ be invariant when $\dd_0$ is rotated by $2\pi/3$. 
These symmetry constraints lead to the following two-parameter lowest-harmonic parametrization:
\begin{equation}
\Delta_{\mathfrak{\ell}}(\dd_0) = 2 V \sum_{j=1, 3, 5} \cos(\GG_j \cdot \dd_0 + \ell \psi),
\label{Potential}
\end{equation}
where $\ell =1$ for the $\mathfrak{b}$ layer and $\ell=-1$ for the $\mathfrak{t}$ layer,
$\GG_j$ is the reciprocal lattice vector obtained by counterclockwise rotation of $\GG_1=(4\pi)/(\sqrt{3}a_0) \hat{y}$ by angle $(j-1)\pi/3$, 
$a_0$ is the monolayer TMD lattice constant, and $V$ and $\psi$ respectively characterize the amplitude and shape of the potentials.
Note that we have chosen the spatial averages of $\Delta_{\mathfrak{b},\mathfrak{t}}$, which must be identical,
as the zero of energy.  

The $\dd_0$ dependence of $\Delta_T$ is most conveniently understood by 
assuming a two-center approximation \cite{Bistritzer2011} for tunneling between 
the metal $d_{x^2-y^2}+id_{xy}$ orbitals, and using a lowest-harmonic approximation.  
This leads to, 
\begin{equation}
\Delta_{T}(\dd_0) = w (1+ e^{-i \GG_2 \cdot \dd_0}+ e^{-i \GG_3 \cdot \dd_0}),
\label{Tunneling}
\end{equation}
where $w$ is a tunneling strength parameter. 
It is informative to highlight three high-symmetry displacement values:
$\dd_{0, n} = n (\ba_1+\ba_2)/3$ for $n =0,\pm 1$, where  
$\ba_{1,2}$ are the primitive translation vectors of the aligned 
bilayer: $\ba_1 = a_0(1,0)$ and $\ba_2 = a_0(1/2,\sqrt{3}/2)$.
For $n=0$ the metal atoms of the two layers are aligned, 
$\Delta_{\mathfrak{t}}=\Delta_{\mathfrak{b}}= 6 V \cos(\psi)  $ and $\Delta_{T} = 3 w$; the valence band maximum states are then 
symmetric and antisymmetric combinations of the isolated layer states.
For $n=\pm 1$ the metal atoms in one layer are aligned with the chalcogen atoms in the other layer, 
and $\Delta_{T}$ vanishes as a result of the threefold rotational symmetry $\hat{C}_{3z}$.
We determine the model parameters by fitting the eigenvalues of $\mathcal{H}_{\uparrow}(\kk=0)$ at the three displacements
to corresponding values from fully relativistic band structure 
calculations using Quantum Espresso \cite{giannozzi2009}.
We find that $(V, \psi, w) \approx $ (8 meV, $-89.6^{\circ}$, $-8.5$ meV) for MoTe$_2$.

\begin{figure}[t]
    \includegraphics[width=1\columnwidth]{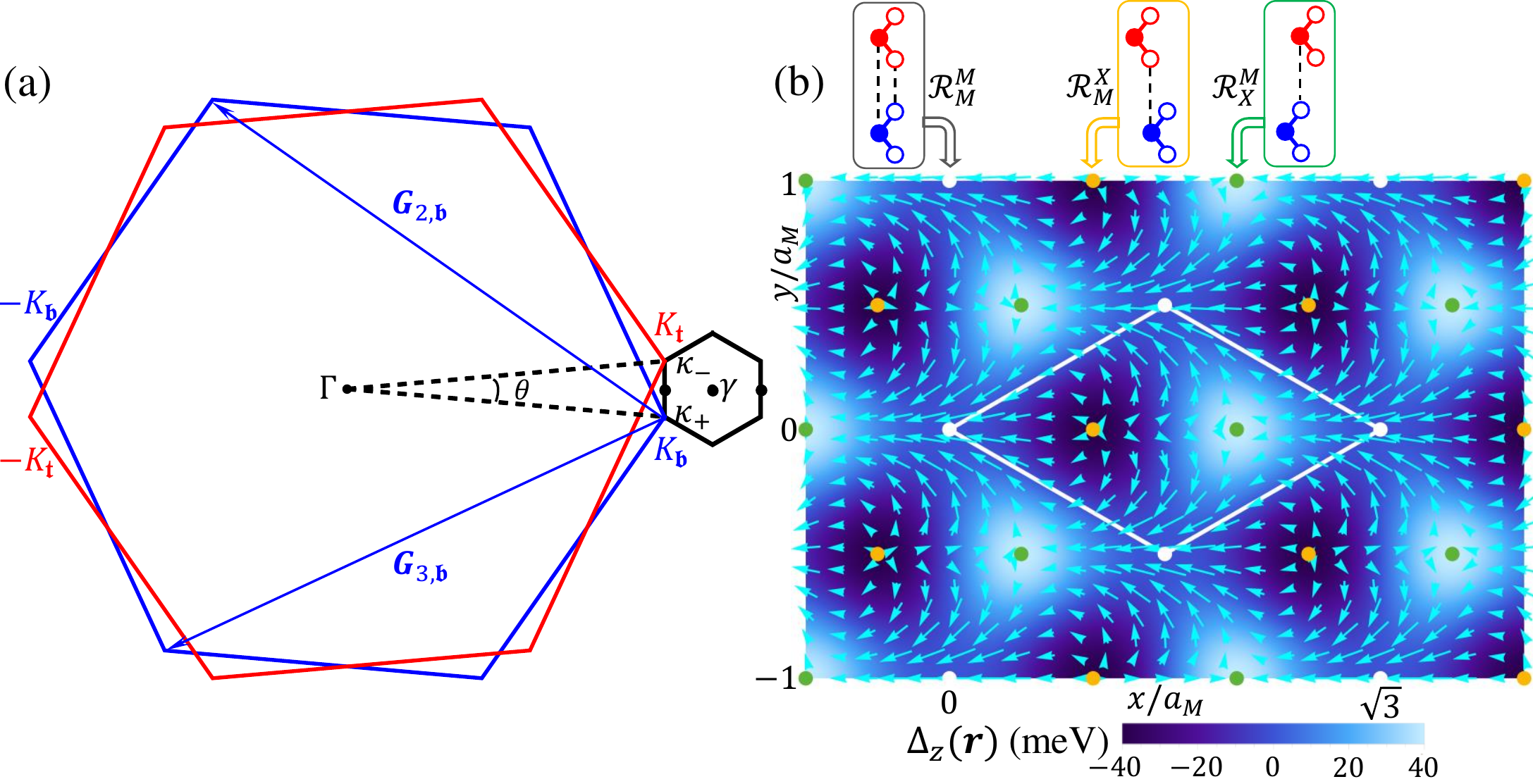}
	\caption{(a) Brillouin zones of the bottom (blue) and top (red) layers in a twisted bilayer, and the moir\'e Brillouin zone (black). (b) The $+K$-valley layer pseudospin skyrmion lattice in the moir\'e pattern. The color map illustrates the variation of $\Delta_z$, and the arrows indicate $\Delta_{x, y}$.
	The white lines outline a single moir\'e unit cell. The dots indicate the high symmetry positions $\mathcal{R}_{M}^{M}$, $\mathcal{R}_{X}^{M}$
	and $\mathcal{R}_{M}^{X}$, where the local interlayer displacements are respectively $\dd_{0,0}$, $\dd_{0,1}$ and $\dd_{0,-1}$.}
	\label{Fig:Skyrmion}
\end{figure}

{\it Moir\'e Hamiltonian.---}
We construct the twisted bilayer Hamiltonian by starting from an aligned bilayer with $\dd_0=0$
and then rotating the bottom and top layers by angles $-\theta/2$ and $+\theta/2$ 
around a metal site.  (Any initial displacement just shifts the moir\'e pattern globally \cite{Bistritzer2011, Wu2018}.)
We take the origin of coordinates to be on this rotation axis and  midway between layers.
With respect to this origin, the bilayer has $D_3$ point group symmetry generated by the threefold 
rotation $\hat{C}_{3z}$ around $\hat{z}$ axis 
and a twofold rotation $\hat{C}_{2y}$ around $\hat{y}$ axis that swaps the two layers. 
In a long-period moir\'e pattern, the local displacement between the two layers, approximated by $\theta \hat{z}\times \rr$, varies smoothly with the 
spatial position $\rr$ \cite{Jung2014, Wu2017}. 
The moir\'e Hamiltonian is
\begin{equation}
\mathcal{H}_{\uparrow}=\begin{pmatrix}
-\frac{\hbar^2 (\kk-\boldsymbol{\kappa}_+)^2}{2 m^*} +\Delta_{\mathfrak{b}}(\rr) & \Delta_{T}(\rr) \\
\Delta_{T}^{\dagger}(\rr) & -\frac{\hbar^2 (\kk-\boldsymbol{\kappa}_-)^2}{2 m^*} +\Delta_{\mathfrak{t}}(\rr)
\end{pmatrix},
\label{HMoire}
\end{equation}
where $\Delta_{\alpha}(\rr)$ is obtained by replacing $\dd_0$ in Eqs.~(\ref{Potential})-(\ref{Tunneling}) with $\theta \hat{z}\times \rr$
to account for the spatial variation of the local inter-layer coordination.  The moir\'e Hamiltonian 
is periodic with the moir\'e period $a_M = a_0/\theta$.  
Because of the twist, the $+K$ points associated with the bottom and top layers are rotated to 
different momenta, accounted for by the $\boldsymbol{\kappa}_{\pm}$ shifts in (\ref{HMoire}).  
We choose a moir\'e Brillouin zone (MBZ)  in which the $\boldsymbol{\kappa}_{\pm}$ points 
are located at the MBZ corners, as illustrated in 
Fig.~\ref{Fig:Skyrmion}(a).

To reveal the spatial structure of the $\Delta_{\alpha}$ field, 
we define the layer pseudospin magnetic field:
\begin{equation}
\boldsymbol{\Delta}(\rr)=(\Delta_x, \Delta_y, \Delta_z) \equiv (\text{Re} \Delta_{T}^{\dagger},\text{Im} \Delta_{T}^{\dagger} , \frac{\Delta_{\mathfrak{b}}-\Delta_{\mathfrak{t}}}{2}).
\end{equation}
As illustrated in Fig.~\ref{Fig:Skyrmion}(b), $\Delta_z(\rr)$ vanishes along the links that connect nearest-neighbor $\mathcal{R}_{M}^{M}$ sites and has minimum 
and maximum values at $\mathcal{R}_{M}^{X}$ and $\mathcal{R}_{X}^{M}$.  
The in-plane pseudospin field, which accounts for 
interlayer tunneling, has vortex and antivortex structures centered on $\mathcal{R}_{M}^{X}$ and $\mathcal{R}_{X}^{M}$. 
Here $\mathcal{R}_{\alpha}^{\beta}$ denotes high-symmetry sites at which $\alpha$ atoms of the bottom layer are locally aligned with $\beta$ atoms of the top layer.  
It follows that $\boldsymbol{\Delta}(\rr)$ forms a skyrmion lattice,
{\it i.e.}, that the direction of the $\boldsymbol{\Delta}(\rr)$ covers the unit sphere once 
in each moir\'e unit cell (MUC).  We have explicitly confirmed this property by 
numerically evaluating the winding number \cite{Moon1995}:
\begin{equation}
N_w \equiv \frac{1}{4\pi}\int_{\text{MUC}} d\rr \frac{\boldsymbol{\Delta} \cdot(\partial_x \boldsymbol{\Delta} \times \partial_y \boldsymbol{\Delta} )}{|\boldsymbol{\Delta}|^3} =-1.
\end{equation}
Skyrmion lattice pseudospin textures in position space 
indicate \cite{Nagaosa2013} the possibility of topological electronic bands in momentum space,
although we will find that the connection is not one-to-one.

\begin{figure}[t!]
    \includegraphics[width=1\columnwidth]{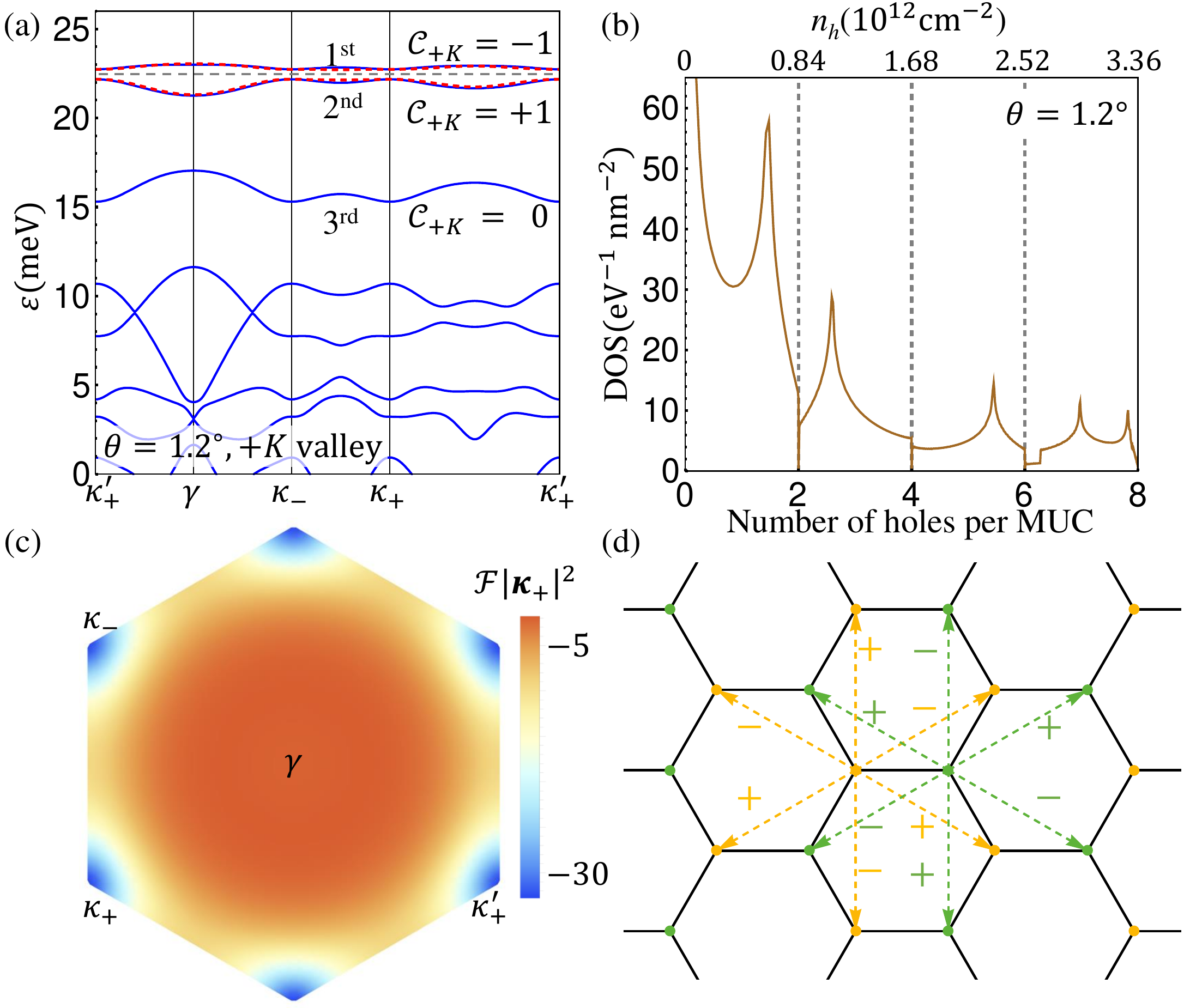}
	\caption{(a) Moir\'e band structure at twist angle $1.2^{\circ}$. The system is a topological insulator when the chemical potential (black dashed line) is in the gap between the first and the second bands. 
	The red dashed lines are a tight-binding-model fit based on the effective Hamiltonian (\ref{HKM}) with $t_0 \approx 0.29 $ meV and $t_1 \approx 0.06 $ meV. 
	(b) Total density of states (DOS) as a function of the number of holes per moir\'e unit cell (bottom) and per area (top). (c) Berry curvature $\mathcal{F}$ for the first band in (a). Here the typical magnitude of $\mathcal{F}$ is three orders of magnitude larger than that in the monolayer \cite{Di2012, SM}. (d) Illustration of the tight-binding model (\ref{HKM}).
	The yellow and green dots represent $\mathcal{R}_{M}^X$ and $\mathcal{R}_{X}^M$ sites, and together form a honeycomb lattice. The signs $\pm$ refer to the bond and spin dependent hopping phase factors $\exp(\pm i  2\pi s/3)$.}
	\label{Fig:Band}
\end{figure}

{\it Topological bands.---}
The moir\'e band structure is illustrated in Fig.~\ref{Fig:Band}(a) for a representative angle $\theta=1.2^{\circ}$ .
The $\hat{C}_{2y} \hat{\mathcal{T}}$ symmetry of the Hamiltonian 
maps $\boldsymbol{\kappa}_+ \to \boldsymbol{\kappa}_-$ and therefore enforces degeneracy between
these points.
For the two topmost moir\'e bands of the $+K$ valley, wave functions in
the $\mathfrak{b}$ ($\mathfrak{t}$) layer are concentrated near the 
$\mathcal{R}_{X}^{M}$ ($\mathcal{R}_{M}^{X}$) sites, 
which are $\Delta_{\mathfrak{b}}$ ($\Delta_{\mathfrak{t}}$) maxima. 
Because of the layer-dependent momentum shifts $\boldsymbol{\kappa}_{\pm}$ in the kinetic energies,
the moir\'e band wave functions vary rapidly over the MBZ.  
In particular, the wave function of the topmost moir\'e band at $\boldsymbol{\kappa}_+$ 
and $\boldsymbol{\kappa}_-$ are respectively localized in layers $\mathfrak{b}$ and $\mathfrak{t}$.
By integrating the Berry curvature 
$\mathcal{F}$ over the MBZ \cite{Xiao_RMP}, we confirm that the 
Chern numbers $\mathcal{C}$ of the two topmost $+K$ valley moir\'e bands in Fig.~\ref{Fig:Band}
are non-trivial ($\mathcal{C}=\pm 1$) at $\theta=1.2^{\circ}$.
The corresponding bands at the $-K$ valley must have the opposite Chern numbers due to the $\hat{\mathcal{T}}$ symmetry. 
Spin-valley locking implies that when the chemical potential is in the gap between the two topmost bands, 
the twisted homobilayer is not only a valley Hall insulator but also a quantum spin Hall insulator, {\it i.e.}, a topological insulator \cite{KaneMele2005, KaneMeleZ2}.

To gain deeper insight into the topological bands, we construct a tight binding model.
The real space distribution of the wave functions suggests a two-orbital model for the first two moir\'e bands:
\begin{equation}
\begin{aligned}
H_{\text{TB}} & = \sum_{\ell, s} \sum_{\RR \RR' }^{'} t_0 \, c_{\RR \ell s}^{\dagger} c_{\RR' (-\ell) s}\\
 & +\sum_{\ell, s} \sum_{\RR} \sum_{\ba_M}^{'} t_1 e^{ i s \boldsymbol{\kappa}_\ell \cdot \ba_M} \, c_{(\RR+\ba_M) \ell s}^{\dagger} c_{\RR \ell s},
\end{aligned}
\label{HKM}
\end{equation}
where $s = \pm$ denotes spin (equivalent to valley $\pm K$), and
$\ell = \pm $ labels orbitals localized in the bottom (+1) and top ($-1$) layers
and centered around the $\mathcal{R}_{X}^{M}$ and $\mathcal{R}_{M}^{X}$ sites.
The two orbitals form a honeycomb lattice in Fig.~\ref{Fig:Band}(d).
In (\ref{HKM}), the spin up and down sectors are decoupled due to the spin-valley $U(1)$ symmetry of the low-energy theory, and
are related by $\hat{\mathcal{T}}$ symmetry. The first line of (\ref{HKM}) captures 
inter-layer hopping between nearest neighbors on the honeycomb lattice.
Its form is constrained by the requirements that the energy spectra have threefold rotational symmetry
and be identical at $\kappa_+$ and $\kappa_-$ points. 
The second line of (\ref{HKM}) captures intralayer 
hopping between next nearest neighbors on the honeycomb lattice; 
the bond and spin-dependent  
phase factors $\exp ( i s \boldsymbol{\kappa}_\ell \cdot \ba_M)$, which take values of $\exp (\pm i 2\pi/3)$, are analogous to the Peierls substitution and account for the momentum 
shift $\boldsymbol{\mathcal{\kappa}}_\ell$ in (\ref{HMoire}).
The Hamiltonian (\ref{HKM}) is equivalent to the 
Kane-Mele model \cite{KaneMele2005, KaneMeleZ2}, and to two time-reversed-partner 
copies of the Haldane model \cite{Haldane1988}.  
It correctly captures both the topological character and the energy 
dispersion of the first two bands in Fig.~\ref{Fig:Band}(a).

\begin{figure}[t]
    \includegraphics[width=1\columnwidth]{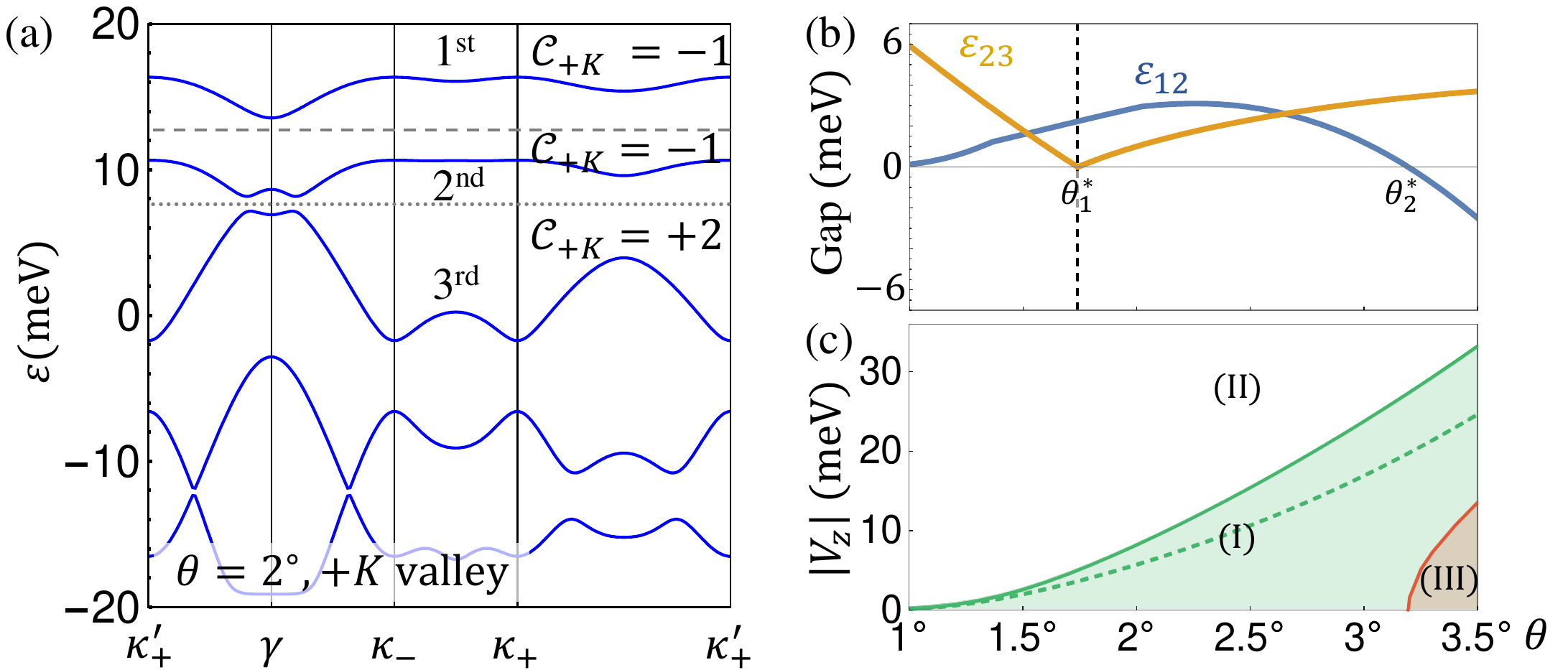}
	\caption{(a) Moir\'e bands at twist angle $2^{\circ}$. (b) Energy gaps between adjacent moir\'e bands as a function of $\theta$. The gap $\varepsilon_{ij}$ is the difference between the minimum energy of band $i$ and the maximum energy of band $j$. (c) Phase diagram as a function of angle $\theta$ and layer bias potential $V_z$. When the chemical potential is in the gap between the first and the second band, the system is a topological insulator in region (I) and a trivial insulator in region (II). In region (III), the two bands overlap in energy. The solid and dashed green lines show the critical bias potentials obtained using the full moir\'e Hamiltonian and the effective tight-binding model respectively. (b) and (c) have the same horizontal axis. }
	\label{Fig:Pdiagram}
\end{figure}

Beyond a critical angle $\theta_1^* \approx 1.74^{\circ}$
the gap between the second and the third bands closes at the $\gamma$ point, as illustrated in Fig.~\ref{Fig:Pdiagram}.  
When $\theta$ crosses $\theta_1^*$ from below, the Chern number of the first  band in $+K$ valley
remains as $-1$, while
the Chern numbers for the second and third bands change from $(+1, 0)$ to $(-1, +2)$. 
Although, the two-orbital model (\ref{HKM}) is not fully applicable for $\theta > \theta_1^*$ 
it still captures the main character of the first two bands in 
regions of momentum space away from the $\gamma$ point.
The system remains as a topological insulator when the chemical 
potential is in the gap between the first and the 
second bands until $\theta$ reaches $\theta_2^* \approx  3.1^{\circ}$, 
beyond which there is no global gap between the first two bands. In the SM\cite{SM}, we have verified the robustness of our predicted topological bands against perturbation from remote bands.

{\it Field induced topological transition.---}
Because the two sublattices in (\ref{HKM}) are associated with different layers, 
a vertical electric field generates a staggered sublattice potential, which can induce a topological phase transition \cite{Haldane1988, KaneMele2005, KaneMeleZ2}. To study this transition, we add a layer dependent potential $\ell V_z/2$ to the moir\'e Hamiltonian (\ref{HMoire}) so 
that $\Delta_z \to (\Delta_{\mathfrak{b}}-\Delta_{\mathfrak{t}}+V_z)/2$.  (We neglect the small spatial modulation 
of $V_z$ due to variation in the vertical distance between layers in the moir\'e pattern \cite{Zhang_Shih}.)
The magnitude of $V_z$ has a critical value $|V_z|_c$, at which the gap between the first and the second moir\'e bands closes at $\kappa_{\pm}$  points. When $|V_z|>|V_z|_c$, wave functions in the first moir\'e band are primarily localized in 
one single layer and the band becomes topologically trivial.
The tight-binding model (\ref{HKM}) predicts that $|V_z|_c$ is equal to splitting between the first and the second bands 
at  $\kappa_{\pm}$ when $V_z=0$, because the interlayer hopping term in (\ref{HKM}) vanishes at these momenta.  In Fig.~\ref{Fig:Pdiagram}(c) we compare values of 
$|V_z|_c$ calculated from the tight-binding and the full moir\'e band Hamiltonian,
showing that they match well, particularly for small twist angles (long moir\'e period).
We note that there is no one-to-one correspondence between the Chern numbers $\mathcal{C}$ of the electronic bands 
and the winding number $N_w$ of the pseudospin field, which remains non-trivial 
until $V_z$ equals $|\Delta_{\mathfrak{b}}-\Delta_{\mathfrak{t}}|$ evaluated at the
$\mathcal{R}_{M}^{X}$ or $\mathcal{R}_{X}^{M}$ sites.

{\it Interaction effects.---}
When the moir\'e bands are nearly flat, the density of states is strongly enhanced [Fig.~\ref{Fig:Band}(b)] and many-body interaction effects are magnified.
Here we focus on interaction effects within the first two moir\'e bands at zero $V_z$ and small $\theta$.
The on-site Coulomb repulsion $U_0$ scales as $e^2/(\epsilon a_W)$, 
where $\epsilon$ is an effective dielectric constant that depends on the three-dimensional dielectric environment, and $a_W$ is the spatial 
extent of the Wannier orbitals centered at $\mathcal{R}_{M}^X$ or $\mathcal{R}_{X}^M$  sites. 
For $\theta$ around $1^{\circ}$,  we find that $U_0$ can be more than one order of magnitude larger than the 
hopping parameters $t_{0,1}$ \cite{SM}.  
In the strong correlation limit, we anticipate that the interplay between layer and spin/valley degrees of freedom
will lead to unusual distinct insulating states at integer numbers of holes per MUC.
For one hole per MUC, where the first moir\'e band is half filled,
one candidate insulating state is ferromagnetic. Because the single-particle Hamiltonian has only $U(1)$ symmetry,
perpendicular spin polarization is energetically preferred. The Ising spin anisotropy implies
finite temperature phase transitions.  When the first moir\'e band is completely spin-polarized, 
the system is a quantum anomalous Hall insulator.  
Similar physics could occur for three holes per MUC, where the second moir\'e band is half filled.
For two holes per MUC (equivalently one hole per sublattice site of the honeycomb lattice in Kane-Mele model), there is a competition between the quantum spin Hall insulator and the antiferromagnetic insulator \cite{Assaad2011}, which occur 
for weak and strong interactions respectively.
For some fractional numbers of holes per MUC, the flat bands may host fractional topological 
insulators \cite{Qi2011}.

{\it Discussion.---}
It has been proposed that Hubbard model can be simulated in 
TMD {\it heterobilayers} \cite{Wu_Hubbard2018}. In twisted TMD homobilayers, the two layers can be effectively decoupled 
by using a finite layer bias potential to drive the system into region (II) of 
the phase diagram in Fig.~\ref{Fig:Pdiagram}(c).  Thus, conventional one-orbital Hubbard model can 
also be studied in twisted homobilayers, with a greater scope for {\it in situ} manipulation of model parameters.  Compared to heterobilayers, twisted TMD homobilayers may be experimentally realized with a more precise control of the twist angle by using the `tear-and-stack' technique\cite{Kim_bilayer,  Cao2018Magnetic, Cao2018Super}. In the SM \cite{SM}, we also show that a two-orbital Hubbard model can be simulated using twisted TMD homobilayers in the AB stacking configuration.
Moir\'e bands with valley-contrasting Chern numbers have been proposed in some graphene-based moir\'e systems \cite{Song2015, Senthil2018, Jung2018}. In this case 
however,  
quantum spin Hall states that might be induced by interactions cannot survive to accessible temperatures because electrons 
in graphene have accurate $SU(2)$ spin symmetry which enhances fluctuation effects.
In Ref.~\cite{Tong2017}, quantum spin Hall nano-dots and nano-stripes have been proposed 
for TMD-based moir\'e systems in which the large gap between valence and conduction bands needs to be inverted by strong vertical electric field. In contrast, our model Hamiltonian relies only
on valence band states. Our proposal for topological states is based on valley contrast physics and on  pseudospin texture in the moir\'e pattern; the advantage is that it does not require massless chiral fermions in the parent monolayer or aligned bilayer, which may lead to application in a larger class of two-dimensional materials.

F.W. and I.M. are supported by the Department of Energy, Office of Science,
Materials Science and Engineering Division.  F.W. is also supported by  Laboratory
for Physical Sciences.
Work at Austin is supported by Army Research Office (ARO) Grant W911NF-17-1-0312 (MURI) and by the Welch foundation under Grant TBF1473.
We acknowledge HPC resources provided by the Texas Advanced Computing Center (TACC) at The University of Texas at Austin.

\bibliographystyle{apsrev4-1}
\bibliography{refs}

\clearpage
\begin{center}
	\textbf{Supplemental Material}
\end{center}

This Supplemental Material includes the following five sections: (1)DFT band structure, (2) a four-band model with remote conduction bands, (3) a four-band model with remote spin-split valence bands, (4) interaction effects in the Kane-Mele model, and (5) simulation of two-orbtial Hubbard model using the AB stacking configuration where the two layers are relatively rotated by an angle near $\pi$.

\section{DFT band structure}
We perform fully-relativistic density functional theory (DFT) calculations under the local density approximation (LDA)  for aligned MoTe$_2$ bilayers using Quantum Espresso \cite{giannozzi2009}. The Perdew-Zunger (LDA)  exchange-correlation
functional has been used. We choose a plane-wave cutoff energy of 50 Ry and a corresponding charge density cutoff energy of 408 Ry, and we sample the Brillouin zone with a 16$\times$16$\times$1 grid of $k$ points. 
The thickness of
the vacuum layer is about 20 \AA.
The in-plane lattice constant $a_0$ is taken to be 3.472 \AA.\cite{Liu2013,Agarwal1972}  The in-plane atomic positions are fixed, while the out-of-plane  positions have been relaxed for all atoms until the out-of-plane direction force on each atom is less than 0.005 eV/\AA.

\begin{figure}[t!]
	\includegraphics[width=1\columnwidth]{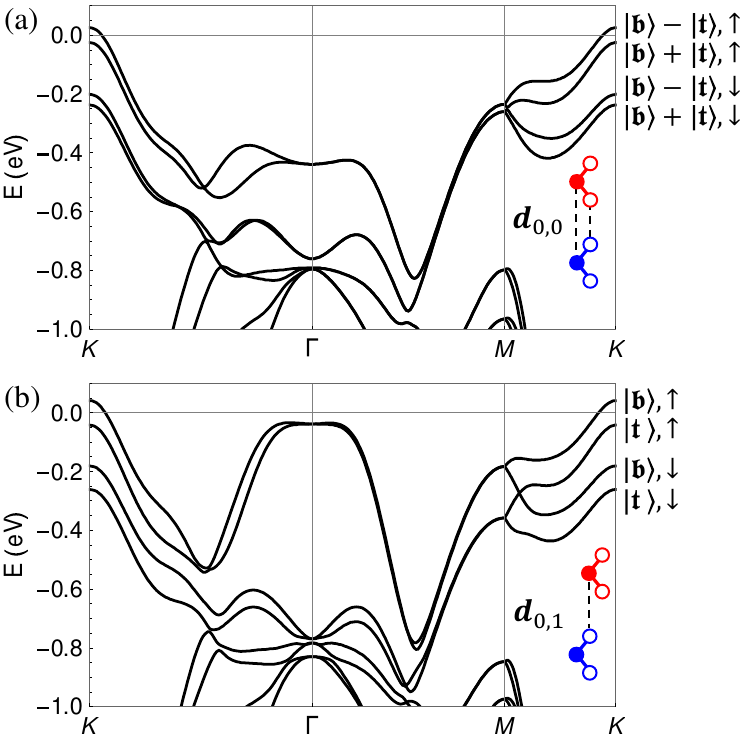}
	\caption{DFT band structure for AA stacked MoTe$_2$ bilayers with $\theta=0$. The in-plane displacement between the two layers are $\dd_{0,0}$ in (a) and $\dd_{0,1}$ in (b). The major layer and spin characters of the valence bands at $+K$ point are listed, which are fully consistent with our analytic model. The zero energy is taken to be the same as that in Eq.~(1) of the main text. }
	\label{Fig:DFT}
\end{figure}

The band structures for AA stacked MoTe$_2$ bilayers ($\theta=0$) are shown in Figs.~\ref{Fig:DFT}(a, b) respectively for in-plane displacements $\dd_{0,0}$ and $\dd_{0,1}$.  We note that bilayers with displacements $\dd_{0,1}$ and $\dd_{0,-1}$ are related by a mirror operation $z \leftrightarrow -z$; therefore the two structures have identical energy dispersion, while the wave function characters are related by the mirror operation. We measure the layer  separation by the vertical distance $D_z$ between the metal (Mo) atoms associated with different layers. With this definition, $D_z$ is respectively 7.8\AA~ and 6.9\AA~ for $\dd_{0, 0}$ and $\dd_{0, 1}$ structures. 

As shown in Fig.~\ref{Fig:DFT}, the valence band maxima are located in  $\pm K$ valleys for both $\dd_{0, 0}$ and $\dd_{0, 1}$ structures. Therefore, in our model study, the topmost moir\'e valence  states are derived from $\pm K$ valleys. This is the reason why we choose MoTe$_2$ bilayers, whereas in other TMDs, for example MoS$_2$ bilayers \cite{Naik2018ultraflat}, the valence band maximum is located in $\Gamma$ valley.   
Because the valence band maximum at the $\Gamma$ point is only lower in energy compared to that at $\pm K$ points by about 80 meV in the $\dd_{0, 1}$ structure of MoTe$_2$,
the DFT calculation may not be accurate enough to precisely determine the energy separation between $\Gamma$ and $\pm K$ valence states. However, the $\Gamma$ and $\pm K$ states are always separated by a large momentum, and therefore, they can be studied separately in a low-energy theory.  

To determine the orbital, layer and spin characters of the bands, we obtain a tight-binding Hamiltonian in a basis of Wannier functions using Wannier90. We project onto $p$ orbitals of Te and $d$ orbitals of Mo without performing maximal localization, and reproduce the DFT band structure with high accuracy.

We have also performed DFT study for AA stacked WSe$_2$ bilayer, of which the valence band maxima are located at $\pm K$ points for $\dd_{0,0}$ structure, but at $\Gamma$ point for $\dd_{0,\pm1}$ structures, according to our calculation. Because of this complicated energy landscape, it should be examined experimentally whether low-energy moir\'e bands in twisted  bilayer WSe$_2$ are derived from $\pm K$ valleys or $\Gamma$ valley. More advanced band structure calculation, such as GW calculation, should also be applied to bilayer WSe$_2$ to get a more accurate estimation of the energy landscape.  The effective model in Eq.(2) of the main text applies equally to $+K$ valley moir\'e valence states in twisted  bilayer WSe$_2$ (other TMDs as well); the corresponding parameter values for WSe$_2$ are $(V, \psi, w) \approx $ (8.9 meV, $91^{\circ}$, $9.7$ meV), and there are similar topological moir\'e bands.

\section{Remote conduction bands}
In monolayer TMDs, the $k.p$ Hamiltonian at $\pm K$ valleys is often described by a massive Dirac model in the basis of conduction and valence band sates \cite{Di2012}. This massive Dirac model is particularly convenient for analyses of the valley dependent optical selection rules. In the main text, we have discarded the conduction band states and used a parabolic band approximation for valence band states. There is a question about whether such an approximation is adequate to capture all important low-energy physics, in particular the moir\'e band structure and the topological character. Below we answer this question in the affirmative by studying a moir\'e Hamiltonian that includes both the valence and conduction bands. 

As in the main text, we study spin up states at $+K$ valley. For aligned bilayers, the $k.p$ Hamiltonian that includes both conduction and valence band states is given by:
\begin{equation}
\mathcal{H}_{\uparrow}(\theta=0, \dd_0)=\begin{pmatrix} 
h_{\mathfrak{b}}(\kk, \dd_0) & T(\dd_0) \\
T^{\dagger}(\dd_0) & h_{\mathfrak{t}}(\kk, \dd_0) 
\end{pmatrix},
\label{4band}
\end{equation}
where $h_{\ell}$ is the massive Dirac Hamiltonian in layer $\ell$, and $T(\dd_0)$ is the interlayer tunneling matrix.  $h_{\ell}$ is given by:
\begin{equation}
h_{\ell}(\kk, \dd_0)=\begin{pmatrix} 
\Delta_g+\Delta_{\ell, c}(\dd_0) & \hbar v_F (k_x-i k_y) \\
\hbar v_F (k_x+i k_y)  &  \Delta_{\ell, v}(\dd_0)
\end{pmatrix}.
\label{MDirac}
\end{equation}
Here the two components refer to the metal $d_{z^2}$ and $d_{x^2-y^2}+i d_{xy}$ orbital states, which carry different angular momenta and represent the main character respectively for conduction ($c$) and valence ($v$) band states. In (\ref{MDirac}),$\Delta_g$ is the average band gap,  $\Delta_{\ell, \xi}(\dd_0)$  is the variation of band $\xi$ extrema energy as a function of $\dd_0$, and $v_F$ is the Fermi velocity. Here $\xi$ is the band index for $c$ and $v$. Similar to Eq.(2) in the main text, $\Delta_{\ell, \xi}$  can be parametrized as:
\begin{equation}
\Delta_{\ell, \xi}(\dd_0) = 2V_{\xi}\sum_{j=1, 3, 5} \cos(\GG_j \cdot \dd_0 + \ell \psi_{\xi}).
\label{Dd0lx}
\end{equation} 

The tunneling matrix $T(\dd_0)$ is derived by using a two-center approximation \cite{Bistritzer2011} and also taking into account the threefold-rotational symmetry at high-symmetry displacements \cite{Yong2017}. $T(\dd_0)$ is parametrized as:
\begin{equation}
\begin{aligned}
T(\dd_0)&=\begin{pmatrix} 
w_c & w_{cv} \\
w_{vc} & w_v
\end{pmatrix}\\
&+
\begin{pmatrix} 
w_c & w_{cv} e^{-i2\pi/3} \\
w_{vc} e^{i2\pi/3} & w_v
\end{pmatrix} e^{-i \GG_2\cdot \dd_0} 
\\
&+
\begin{pmatrix} 
w_c & w_{cv} e^{i2\pi/3} \\
w_{vc} e^{-i2\pi/3} & w_v
\end{pmatrix} e^{-i \GG_3\cdot \dd_0}.
\end{aligned}
\label{Td0}
\end{equation}
The mirror operation ($z \leftrightarrow -z$) leads to the constraints that $w_c$ and $w_v$ are real numbers, and $w_{cv}=w_{vc}^*$.  Note that the interlayer tunneling between valence and conduction band states vanishes at $\dd_0 =0 $.

We determine parameter values by fitting to the first-principles band structure at the three high-symmetry displacements.  For $AA$ stacked MoTe$_2$ bilayer, we obtain that $\Delta_g=1.1$ eV, $v_F=0.4\times 10^6$ m/s, $(V_v, \psi_v, w_v)=$ (8 meV,$-89.6^\circ$, $-8.5$ meV), $(V_c, \psi_c, w_c)=$(5.97 meV, $-87.9^\circ$, $-2$ meV) and $w_{cv}= 15.3$ meV. Here $\psi_v$ and $\psi_c$ are close to $-90^{\circ}$, which implies that the scalar potential $(\Delta_{\mathfrak{b}, \xi}+\Delta_{\mathfrak{t}, \xi})/2$ has an amplitude much smaller than that of the potential difference $(\Delta_{\mathfrak{b}, \xi}-\Delta_{\mathfrak{t}, \xi})/2$.  We note that density-functional theory underestimates the band gap $\Delta_g$, but the exact numerical values should not influence qualitative effects that we discuss.  

The band gap $\Delta_g$ is much larger than other energy scales. Therefore, we can integrate out conduction band states using a second-order perturbation theory, and the resulting model is exactly that in Eq. (1) of the main text. The effective mass $m^*$ is then approximated by $\Delta_g/(2 v_F^2)$, and the spatial modulation of $m^*$ is of order $V_\xi/\Delta_g \ll 1 $ and can be neglected.
In the main text, $m^*=0.62 m_e$ has been used where $m_e$ is the electron bare mass.
This provides the first justification that the model with only valence band states is valid.

\begin{figure}[t!]
	\includegraphics[width=1\columnwidth]{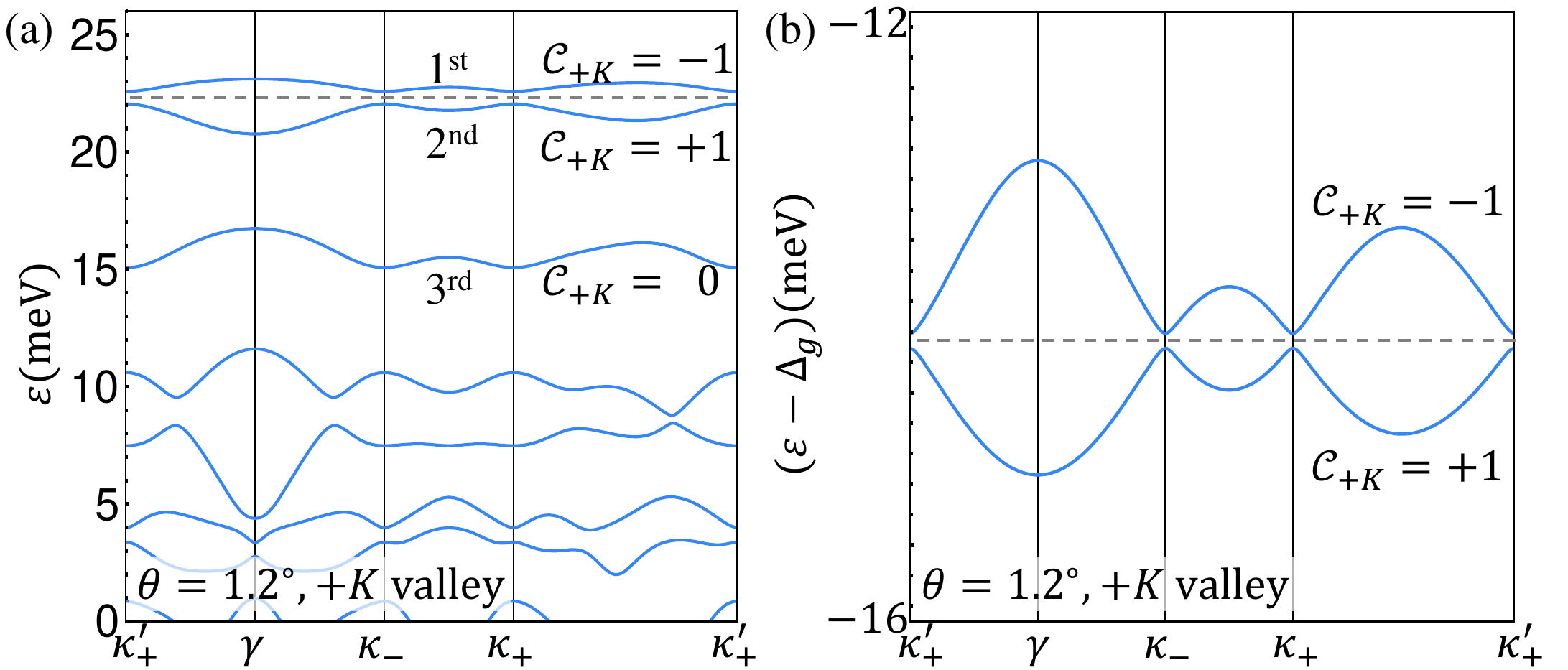}
	\caption{Moir\'e band structure obtained from the four-band model in (\ref{Hmoire4}). The twist angle is $1.2^{\circ}$. (a) Moir\'e valence bands. (b) Two lowest energy moir\'e conduction bands in spin up sector at $+K$ valley. Note that the energy is measured relative to $\Delta_g$ in (b).}
	\label{Fig:DBand}
\end{figure}

We continue with the four-band Hamiltonian (\ref{4band}). When there is a twist between the two layers, the moir\'e Hamiltonian is:
\begin{equation}
\mathcal{H}_{\uparrow}=\begin{pmatrix} 
h_{\mathfrak{b}}(\kk, \rr) & T(\rr) \\
T^{\dagger}(\rr) & h_{\mathfrak{t}}(\kk, \rr) 
\end{pmatrix},
\label{Hmoire4}
\end{equation}
where $\rr$ is the two-dimensional position operator. $h_{\ell}$ in the moir\'e pattern has the form
\begin{equation}
\begin{aligned}
h_{\ell}(\kk, \rr)&=\begin{pmatrix} 
\Delta_g+\Delta_{\ell, c}(\rr) & 0 \\
0  &  \Delta_{\ell, v}(\rr)
\end{pmatrix}\\
& + e^{-i\ell \frac{\theta}{4} \xi_z }[\hbar v_F (\kk-\boldsymbol{\kappa}_{\ell})\cdot \boldsymbol{\xi}]e^{+i\ell \frac{\theta}{4} \xi_z},
\end{aligned}
\end{equation}
where $\boldsymbol{\xi}$ is the Pauli matrix in the band basis. The moir\'e potentials $\Delta_{\ell, \xi}(\rr)$ and the tunneling term $T(\rr)$ are obtained from  Eqs. (\ref{Dd0lx})-(\ref{Td0}) by replacing $\dd_0$ with $\theta \hat{z}\times \rr$. The moir\'e Hamiltonian (\ref{Hmoire4}) is formally similar to that for twisted bilayer graphene \cite{Bistritzer2011,Wu_phonon2018}, and the main differences are the Dirac mass $\Delta_g$.

We diagonalize the moir\'e Hamiltonian (\ref{Hmoire4}) by using a plane-wave expansion and parameter values of MoTe$_2$ bilayer. The moir\'e bands and the Chern numbers are summarized in Fig.~\ref{Fig:DBand}. The moir\'e valence bands and the Chern numbers are consistent with those obtained using the two-band model of the main text. Fig.~\ref{Fig:DBand}(b) shows the two lowest-energy moir\'e conduction bands in spin up sector at $+K$ valley, which  carry Chern numbers $\pm 1$.  The topological character of the moir\'e conduction bands can be understood in the same way as that  for valence bands. 

In monolayer TMD, the valence band edges at valley $\pm K$ already  carry a finite Berry curvature due to the absence of inversion symmetry \cite{Di2012}, and the corresponding Berry curvature at $+K$ point is given by $-2(\hbar v_F/\Delta_g)^2$, which is about $-$0.113 nm$^{2} $ for monolayer MoTe$_2$. In contrast, the typical value of the Berry curvature as shown in Fig.~3 of the main text for the topological moir\'e band is about $-10/|\boldsymbol{\kappa_+}|^2 \approx -156.6 $ nm$^2$, three orders of magnitude larger than the intrinsic monolayer value. We can define a dimensionless quantity $2(\hbar v_F |\boldsymbol{\kappa_+}| /\Delta_g)^2 \sim 7\times 10^{-3}$, which is much smaller than 1 and justifies the two-band model with only  valence band sates.

\begin{figure}[t!]
	\includegraphics[width=1\columnwidth]{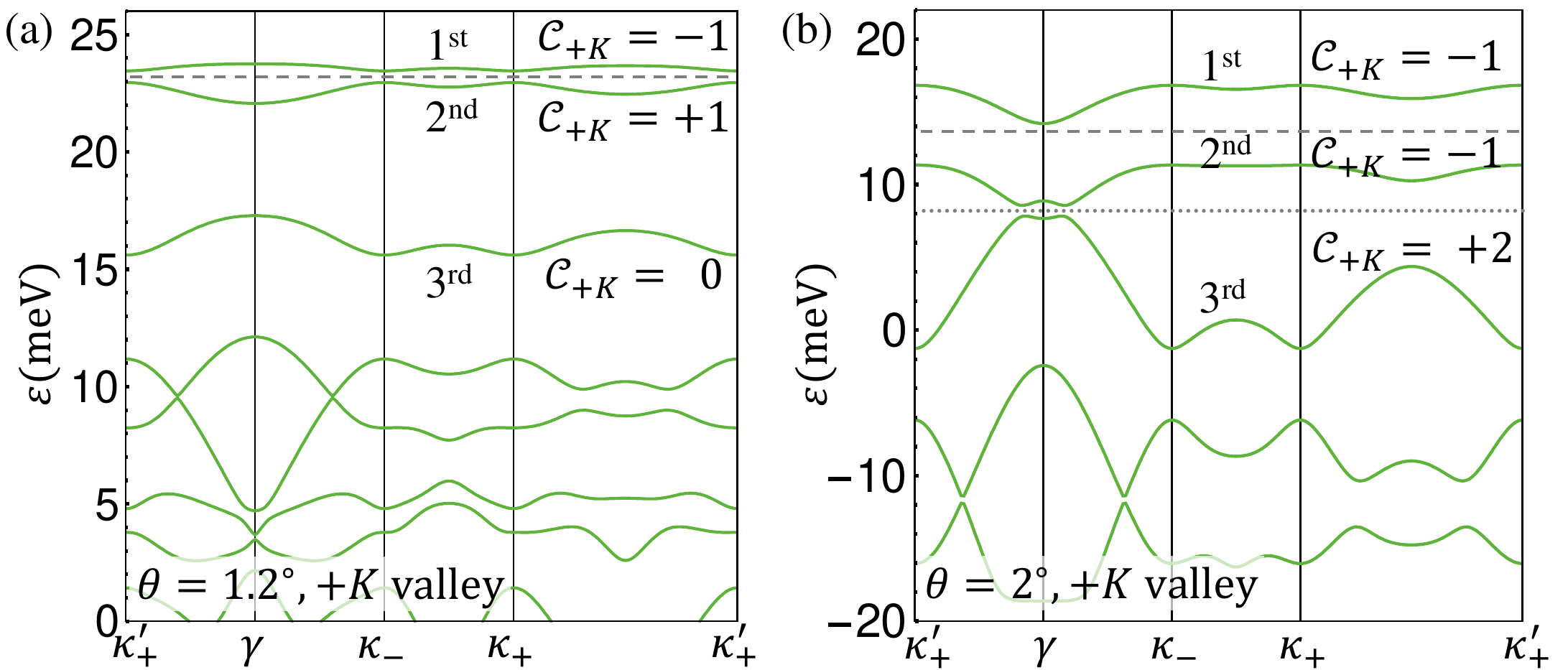}
	\caption{Moir\'e band structure obtained from the four-band model in (\ref{Smixing}) that includes spin mixing effects. The twist angle is $1.2^{\circ}$ in (a) and $2^{\circ}$ in (b). The bands shown in these figures are mainly of spin up character, while spin down bands are separated from them by $\Delta_{SOC} \sim 220 $ meV. }
	\label{Fig:SBand}
\end{figure}

\section{Remote valence bands}
We study spin mixing effects from remote valence bands. In TMDs, the topmost spin up and down valence bands at $\pm K$ points are separated in energy by more than 100 meV.  In monolayer TMDs, the  $ z \leftrightarrow -z$ mirror symmetry can be used to classify the bands into two mirror sectors, and the spin-mixing effects are typically ignored. In bilayer TMDs, this mirror symmetry is generally broken.  

Here we study valence bands at $+K$ valley, including spin-orbit split bands. For aligned bilayers, the $k.p$ Hamiltonian is given by:
\begin{equation}
\mathcal{H}(\theta=0, \dd_0)=\begin{pmatrix} 
h_{\mathfrak{b}}(\kk, \dd_0) & T(\dd_0) \\
T^{\dagger}(\dd_0) & h_{\mathfrak{t}}(\kk, \dd_0) 
\end{pmatrix},
\label{Smixing}
\end{equation}
Here $h_{\ell}$ is the intra-layer Hamiltonian given by:
\begin{equation}
h_{\ell} = \begin{pmatrix} 
-\frac{\hbar^2 \kk^2}{2 m^*} +\Delta_{\ell \uparrow}(\dd_0) & 0 \\
0 &  -\Delta_{SOC} -\frac{\hbar^2 \kk^2}{2 m^*} +\Delta_{\ell \downarrow}(\dd_0)
\end{pmatrix},
\label{hs}
\end{equation}
where the two components refer to spin up and down valence states at valley $+K$. Here we use up and down to refer to the major spin character. In (\ref{hs}), the off-diagonal spin-mixing terms are approximated as zero because these terms vanish at the three high-symmetry displacements $\dd_{0,n}$ due to the three-fold rotational symmetry.  $\Delta_{SOC}$ is the spin splitting, and the variation energy $\Delta_{\ell s}$ is parametrized as:
\begin{equation}
\Delta_{\ell s}(\dd_0) =  2V_{s}\sum_{j=1, 3, 5} \cos(\GG_j \cdot \dd_0 + \ell \psi_{s}).
\end{equation}

The interlayer tunneling does include symmetry-allowed spin-mixing terms:
\begin{equation}
\begin{aligned}
T(\dd_0)&=\begin{pmatrix} 
w_{\uparrow} & w_{\uparrow \downarrow} \\
w_{\downarrow \uparrow} & w_{\downarrow}
\end{pmatrix}\\
&+
\begin{pmatrix} 
w_{\uparrow} &w_{\uparrow \downarrow} e^{-i2\pi/3} \\
w_{\downarrow \uparrow} e^{i2\pi/3} & w_{\downarrow}
\end{pmatrix} e^{-i \GG_2\cdot \dd_0} 
\\
&+
\begin{pmatrix} 
w_{\uparrow} & w_{\uparrow \downarrow} e^{i2\pi/3} \\
w_{\downarrow \uparrow} e^{-i2\pi/3} & w_{\downarrow}
\end{pmatrix} e^{-i \GG_3\cdot \dd_0}.
\end{aligned}
\end{equation}
The mirror operation ($z \leftrightarrow -z$), which maps $\dd_0$ to $-\dd_0$, implies that $w_{\uparrow}$ and $w_{\downarrow}$ are real, and $w_{\uparrow \downarrow} = - w_{\downarrow \uparrow}^*$, because the mirror operation acts as $-i s_z$ in the spin space and $\ell_x$ in the layer space. Note that this mirror operation is not a symmetry for a generic $\dd_0$ expect at $\dd_0=\boldsymbol{0}$.

We again determine the parameter values by fitting to the first-principles band structure, and obtain that $\Delta_{SOC}= 220.5$ meV,$(V_{\uparrow}, \psi_{\uparrow}, w_{\uparrow})=$ (8 meV,$-89.6^\circ$, $-8.5$ meV), $(V_{\downarrow}, \psi_{\downarrow}, w_{\downarrow})=$(7.7 meV, $-88.35^\circ$, $-6$ meV) and $w_{\uparrow \downarrow}=-i 5.6$ meV.

We map the Hamiltonian in (\ref{Smixing}) to the moir\'e Hamiltonian at a finite twist angle. The calculated moir\'e band structure are shown in Fig.~\ref{Fig:SBand}, which are nearly identical to that obtained in the main text. This is expected because the spin mixing terms are at least an order of magnitude smaller compared to the spin-orbit splitting $\Delta_{SOC}$.

In conclusion, the two-band model given in Eq. (1) of the main text provides a faithful low-energy description of moir\'e valence bands at small twist angles. We have used quantum spin Hall insulator as a synonym of two-dimensional topological insulator \cite{KaneMeleZ2}, which is a distinct phase without requirement of spin conservation.

In our theory, spin, strictly speaking, is not a good quantum number, but valley is a good quantum number and plays the role of an effective spin. Therefore, the $Z_2$ invariant of the quantum spin Hall insulator and the valley Chern number are directly connected.  We can denote the Chern number for bands below a chemical potential in valley $\tau K$ as $\mathcal{C}_{\tau K}$, where $\tau$ is $+$ or $-$ for the two valleys. Because of time-reversal symmetry, $\mathcal{C}_{+K}=-\mathcal{C}_{-K}$. The $Z_2$ invariant is given by $[(\mathcal{C}_{+K}-\mathcal{C}_{-K})/2] \mod 2  =\mathcal{C}_{+K} \mod 2 $. Based on this definition, the system is in the quantum  spin Hall insulator phase (two-dimensional time-reversal-invariant topological insulator) in region (I) of Fig. 4(c) [main text] when the chemical potential is in the gap between the first and the second moir\'e bands.

\section{Kane-Mele-Hubbard model}
We provide additional discussion of many-body interactions in the topological moir\'e bands. As in the main text, we focus on interaction effects within the first two moir\'e bands with small twist angle, where the tight-binding model $H_{\text{TB}}$ is applicable.

The interaction strength can be tuned by the three-dimensional dielectric environment, and the range of interaction can be controlled by a nearby metallic gate. Here we only consider the on-site repulsive interaction $U_0$ and assume that longer-range interactions are suppressed due to the nearby gate. Effects of long-range Coulomb interaction on the topological bands are also interesting, which we leave to future exploration.
$U_0$ scales as $e^2/(\epsilon a_W)$, where $\epsilon$ is the effective dielectric constant. $a_W$ is the spatial extent of the Wannier orbitals, which can be estimated based on a harmonic oscillator potential approximation of the moir\'e potential $\Delta_{\ell}(\rr)$ near one of its maximum position \cite{Wu_Hubbard2018}. At $\theta = 1.2^{\circ}$, $a_W$ is about 3 nm; for comparison, the corresponding moir\'e period $a_M$ is 16.6 nm. If we take $\epsilon$ to be 10, then $U_0 \approx e^2/(\epsilon a_W) \approx 47$ meV, which is two orders of magnitude larger than the hopping parameter $t_0 \approx 0.29$ meV at $\theta = 1.2^{\circ}$. Therefore, the ratio $U_0/t_0$, which depends on the dielectric environment and the twist angle, can be much greater than one. 

By combining the tight-binding model and the on-site interaction, we arrive at the Kane-Mele-Hubbard model:
\begin{equation}
H =H_{\text{TB}}+U_0 \sum_{\RR \ell} n_{\RR \ell \uparrow} n_{\RR \ell \downarrow}.
\label{KMH}
\end{equation}
The tight-binding model is given by:
\begin{equation}
\begin{aligned}
H_{\text{TB}} &= \sum_{\ell, s} \sum_{\RR \RR' }^{'} t_0 c_{\RR \ell s}^{\dagger} c_{\RR' (-\ell) s}\\
& +\sum_{\ell, s} \sum_{\RR} \sum_{\ba_M}^{'} t_1 e^{ i s \boldsymbol{\kappa}_\ell \cdot \ba_M}  c_{(\RR+\ba_M) \ell s}^{\dagger} c_{\RR \ell s},\\
&= \sum_{\kk} c_{\kk}^{\dagger}[f_0+f_z\sigma_z s_z + f_x\sigma_x + f_y\sigma_y] c_{\kk},
\end{aligned}
\end{equation} 
where $\sigma_{x,y,z}$ and $s_{x,y,z}$ are Pauli matrices in the sublattice and spin spaces, and $f_{0, x, y, z}$ are functions of the momentum $\kk$. The tight binding model has two energy branches $f_0 \pm \sqrt{f_x^2+f_y^2+f_z^2}$ and each branch is doubly degenerate. The term $f_0$ is not important for the topological properties and is absent in the original Kane-Mele model of Refs.~\cite{KaneMele2005, KaneMeleZ2}. At the filling factor where there are two electrons per unit cell (equivalent to one electron per sublattice site), there is a competition between the quantum spin Hall insulator and the antiferromagnetic Mott insulator in the Kane-Mele-Hubbard model \cite{Assaad2011}. Here we focus on a different filling factor where there are three electrons (equivalent to one hole) per unit cell. At this filling factor, the upper two bands are half filled in the non-interacting limit. Because the upper two bands carry spin contrast Chern numbers, the isolated upper band in either spin sector does not have exponentially localized Wannier states. These nearly flat topological bands are similar to Landau levels which are completely flat but have no exponentially localized Wannier states. It is well established that Landau levels with internal degeneracies (for example spin) develop ferromagnetism when they are partially filled with an integer number of electrons per cyclotron orbit. By making an analogy to the Landau level problems, we anticipate that ferromagnetism is likely at the filling factor of one hole per moir\'e unit cell. 

We now use a mean-field theory to demonstrate the Ising anisotropy in the ferromagnetic state.  We perform a Hartree-Fock approximation for the Hubbard interaction on each site:
\begin{equation}
n_{\uparrow}n_{\downarrow}\approx \frac{n_0^2+\boldsymbol{m}^2}{4}-\frac{1}{2}c^{\dagger} (\boldsymbol{m}\cdot\boldsymbol{s})c
\end{equation} 
where $n_0 = \langle n_{\uparrow} + n_{\downarrow}  \rangle$ and $\boldsymbol{m} = \langle c^{\dagger} \boldsymbol{s} c \rangle $. Here  $\langle ...  \rangle$ represents the mean-field average value, and $n_0 = 3/2$. The full mean-field Hamiltonian is:
\begin{equation}
\begin{aligned}
H_{\text{MF}}  & = 2 \mathcal{N} \frac{\boldsymbol{M}^2}{U_0} \\
&+
\sum_{\kk} c_{\kk}^{\dagger}[f_0+f_z\sigma_z s_z + f_x\sigma_x + f_y\sigma_y + \boldsymbol{M}\cdot \boldsymbol{s}] c_{\kk},
\end{aligned}
\end{equation}
where $\boldsymbol{M}= -U_0 \boldsymbol{m}/2 $, and $\mathcal{N}$ is the number of unit cells. $H_{\text{MF}}$ generally has four non-degenerate energy bands. We consider zero temperature, and assume that the  magnetization is strong enough such that the highest energy band is completely empty and other three bands are fully occupied by electrons. The total energy is then given by:
\begin{equation}
\begin{aligned}
\mathcal{E}(\boldsymbol{M})  &= \text{constant}+ 2 \mathcal{N}  \frac{M^2}{U_0} \\
&-\sum_{\kk} \sqrt{M^2+ \boldsymbol{f}^2 + 2|M|\sqrt{\boldsymbol{f}_{\perp}^2+f_z^2 \cos^2 \theta_{\boldsymbol{M}}} },
\end{aligned}
\end{equation}	
where $\boldsymbol{f}^2 = \boldsymbol{f}_{\perp}^2+f_z^2$, $\boldsymbol{f}_{\perp}^2 = f_x^2+f_y^2$, and the vector $\boldsymbol{M}$ is parametrized by the spherical coordinates $ M(\sin\theta_{\boldsymbol{M}} \cos \phi_{\boldsymbol{M}},\sin\theta_{\boldsymbol{M}} \sin \phi_{\boldsymbol{M}},\cos\theta_{\boldsymbol{M}} )$. The saddle point equation $\partial \mathcal{E} / \partial {\boldsymbol{M}} = 0 $ correctly reproduces the self-consistent equation $\boldsymbol{m} = \langle c^{\dagger} \boldsymbol{s} c \rangle $.  The energy functional $\mathcal{E}(\boldsymbol{M})$ is independent of the azimuthal angle $\phi_{\boldsymbol{M}}$, which is a result of the spin $U(1)$ symmetry of the Hamiltonian (\ref{KMH}). $\mathcal{E}(\boldsymbol{M})$  does depend on the polar angle $\theta_{\boldsymbol{M}}$, and is minimized by placing $\boldsymbol{M}$ along $\hat{z}$ axis. Thus, there is an Ising anisotropy that favors the out-of-plane spin polarization. 

\section{AB stacking configuration}
In TMD bilayers with long period moir\'e pattern, there is another distinct stacking configuration that we label as AB \cite{Wu2018}. AA and AB stacking, which are sometimes respectively referred to as $R$ and $H$ stacking, are distinguished by a 180$^{\circ}$ rotation of the top layer. Below we discuss $\pm K$ valley moir\'e bands in the AB stacking configuration.

\begin{figure}[t]
	\includegraphics[width=1\columnwidth]{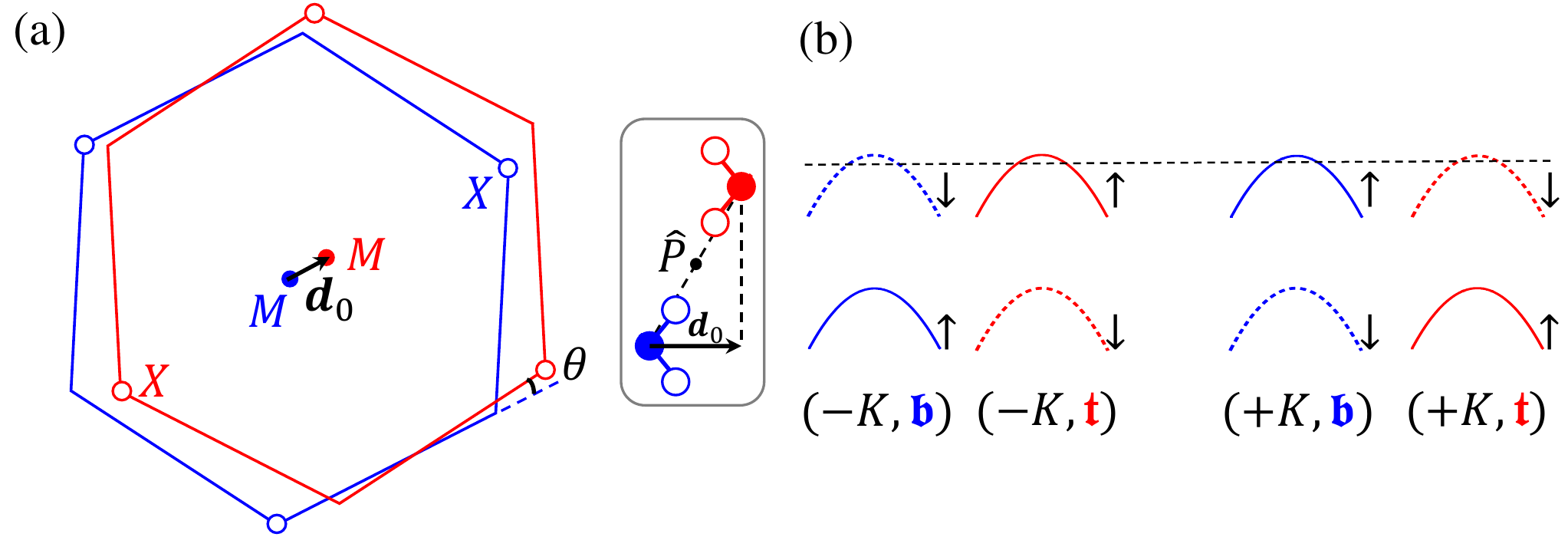}
	\caption{(a) AB stacked TMD $MX_2$ homobilayer. The inset schematically shows the side view at zero twist angle, where the black dot on the dashed line is an inversion center. (b) Schematic plot of the valence bands at $\pm K$ valleys in the AB stacked homobilayer. }
	\label{Fig:Moire_Lattice_AB}
\end{figure}

In AB stacking with zero twist angle and an arbitrary in-plane displacement $\dd_0$, there is always an inversion symmetry $\hat{P}$, and the inversion center is at the midpoint between the two shifted metal atoms located in different layers \cite{Yong2017}, as illustrated in Fig.~\ref{Fig:Moire_Lattice_AB}. The combined symmetry $\hat{P} \hat{\mathcal{T}}$ enforces 
a double degeneracy of the band structure at every momentum. We again focus on $+K$ valley, and electronic properties at $-K$ valley can be inferred by using the $\hat{\mathcal{T}}$ symmetry. At $+K$ valley, the topmost valence bands associated with the bottom and top layer respectively have spin up and down characters, and they are degenerate due to the $\hat{P} \hat{\mathcal{T}}$ symmetry, leading to the following $k.p$ Hamiltonian:
\begin{equation}
\mathcal{H}_{+K}(\theta=0, \dd_0)=\begin{pmatrix}
-\frac{\hbar^2 \kk^2}{2 m^*} +\Delta(\dd_0) & 0 \\
0 & -\frac{\hbar^2 \kk^2}{2 m^*} +\Delta(\dd_0)
\end{pmatrix},
\label{Hd0AB}
\end{equation}
where the two components refer to bottom (spin up) and top (spin down) layers, $\Delta(\dd_0)$ is the variation of the valence band maximum energy as a function of $\dd_0$, and the $2\times 2$ Hamiltonian is proportional to an identity matrix in the layer pseudospin space because of the the $\hat{P} \hat{\mathcal{T}}$ symmetry. In this low-energy Hamiltonian (\ref{Hd0AB}), the two layers are effectively decoupled.

In twisted bilayer, we obtain the moir\'e Hamiltonian by replacing $\dd_0$ with $\theta \hat{z}\times \rr$ and taking into account the momentum shift between the two layers:
\begin{equation}
\mathcal{H}_{+K}=\begin{pmatrix}
-\frac{\hbar^2 (\kk-\boldsymbol{\kappa}_+)^2}{2 m^*} +\Delta(\rr) & 0 \\
0 & -\frac{\hbar^2 (\kk-\boldsymbol{\kappa}_-)^2}{2 m^*} +\Delta(\rr)
\end{pmatrix},
\label{HMoireAB}
\end{equation}
where the two layers are again effectively decoupled and can be studied separately in the single particle Hamiltonian, and the momentum shift $\boldsymbol{\kappa}_{\pm}$ can therefore be removed by a layer-dependent gauge transformation. In each layer electrons  move in the layer-independent potential $\Delta(\rr)$ and form moir\'e bands that are topologically trivial. The topmost moir\'e valence bands can be used to simulate the Hubbard model on a triangular lattice with two orbitals that are centered in different layers but at the same in-plane positions. This two-orbital Hubbard model has an approximate $SU(4)$ symmetry, which would become exact if the vertical separation between the two orbitals were neglected. We will report a detailed study of this two-orbital Hubbard model in a separate work.  A vertical electric field generates an energy difference between the two orbitals, and can lead to the realization of the conventional single-orbital Hubbard model when the field is strong enough  \cite{Wu_Hubbard2018}.     
\end{document}